# Antibunching as a manifestation of light detection back action

M.V. Suslov[1], M.V. Lebedev[1,2*]

[1]Moscow Institute of Physics and Technology (State University)
Dolgoprudny 141701, Moscow region, Russia

[2]Institute of Solid State Physics, Russian Academy of Sciences
Chernogolovka 142432, Moscow region, Russia

[*]Corresponding author e-mail: lebedev@issp.ac.ru

We show that the commonly accepted treatment of the photon antibunching effect as a natural consequence of a probability distribution of particles in a particle flow contradicts the high visibility of the experimentally observed intensity correlation function.

Back action of the measurement process on the measured system is a common place of quantum mechanics. This holds strictly speaking for light emission and detection as well. One cannot consider the light source and the detector as independent because the detection of a photon may change the probability to detect the next one. Being in principle absolutely correct this statement is thought to hold in very special cases only, because every light source consists usually of a macroscopic number of emitting atoms and its quantum nature should be totally obscured by interactions between them and also with their environment. These considerations give a possibility for a semi-classical theory in which the detection process affects to the best end the emitted light field but not the emission itself.

On the other hand the existence of a back action of the measuring process on the studied system may serve as a proof of its quantum nature, and may help to distinguish between fundamental quantum mechanical and nonlinear classical feedback effects. This idea is used in the well known Legget – Garg inequalities [1]. To test these inequalities one has to perform three subsequent measurements on a studied system and repeat this set of measurements for identical prepared systems many times to get correlators $Q_{12}$, $Q_{23}$, and $Q_{13}$ between outcomes of measurements 1-2, 2-3 and 1-3.

We show in this paper that in some cases to distinguish between quantum and classical behavior is possible when measuring the probability of a positive outcome of measurement 2 made with a variable time delay after a positive outcome of measurement 1.

Consider a stationary random source which emits particles and a detector which gives a standard short electric pulse when detecting a particle. Suppose the detector is not ideal and detects a particle with a probability $p$ and overlooks it with a probability $1-p$. Suppose also that the duration of the electric pulse is extremely short and the flux of the particles is rather low so one can neglect the events when two or more particles arrive at the detector simultaneously. If one has a clock which is started at the moment of a particle detection and stopped at the moment of a detection of the next particle and a counter which counts such events with a given time interval $t$ between them one can measure a probability distribution density $F(t)$ which represents the probability $F(t)dt$ to find two subsequent particles in a flux separated with a time interval $t$ with time resolution $dt$. This probability distribution density always exists while it is measured experimentally. In a case of classical theory the measurement process does not affect the emission of particles. This implies the existence of a probability distribution density $f(t)$ for the source to emit two subsequent particles separated with a time interval $t$. In quantum mechanics such a probability distribution density may not exist, because every measurement changes the common state of the source and the emitted particles and affects the probability of the next particle detection.

Let us start with classical theory to make the manifestation of possible quantum mechanical effects more clear. Suppose the stationary random source emits classical particles, e. g. tennis bolls. Then

$$F(t) = pf(t) + p(1-p)\int_{-\infty}^{+\infty} f(t-t_1)f(t_1)dt_1 + \cdots \qquad (1)$$

$$+ p(1-p)^k \iint_{-\infty}^{+\infty} f(t-t_k)f(t_k - t_{k-1})\ldots f(t_1)dt_k \ldots dt_1 + \cdots$$

Where

$$f(t) \geq 0, \ f(t) = 0 \text{ for } t < 0 \text{ and } \int_{-\infty}^{+\infty} f(t)dt = 1 \qquad (2)$$

Is the probability density for the source to emit two subsequent particles separated with a time interval $t$.

Let $\varphi(\omega)$ be a Fourier transform of $f(t)$ such as

$$f(t) = \int_{-\infty}^{+\infty}\frac{d\omega}{2\pi}\varphi(\omega)e^{i\omega t} \text{ and } \varphi(\omega) = \int_{-\infty}^{+\infty} f(t)e^{-i\omega t}dt \qquad (3)$$

Then

$$F(t) = p\int_{-\infty}^{+\infty}\frac{d\omega}{2\pi}\varphi(\omega)e^{i\omega t} + \cdots + p(1-p)^k \iint_{-\infty}^{+\infty} dt_k \ldots dt_1 \int_{-\infty}^{+\infty}\frac{d\omega_k}{2\pi}\varphi(\omega_k)e^{i\omega_k(t-t_k)} \times$$

$$\int_{-\infty}^{+\infty}\frac{d\omega_{k-1}}{2\pi}\varphi(\omega_{k-1})e^{i\omega_{k-1}(t_k-t_{k-1})} \ldots \int_{-\infty}^{+\infty}\frac{d\omega_1}{2\pi}\varphi(\omega_1)e^{i\omega_1(t_2-t_1)} \int_{-\infty}^{+\infty}\frac{d\omega}{2\pi}\varphi(\omega)e^{i\omega t} + \cdots$$

$$= p\int_{-\infty}^{+\infty}\frac{d\omega}{2\pi}\varphi(\omega)e^{i\omega t} + \cdots + p(1-p)^k \int_{-\infty}^{+\infty}\frac{\prod_{i=0}^{k}d\omega_i}{2\pi}e^{i\omega_k t}\delta(\omega-\omega_1)\delta(\omega_1-\omega_2)\ldots\delta(\omega_k-\omega_{k-1}) \times$$

$$\prod_{i=0}^{k}\varphi(\omega_i) + \cdots = p\int_{-\infty}^{+\infty}\frac{d\omega}{2\pi}\varphi(\omega)e^{i\omega t} + \cdots + p(1-p)^k \int_{-\infty}^{+\infty}\frac{d\omega}{2\pi}\varphi(\omega)^{k+1}e^{i\omega t} + \cdots \qquad (4)$$

Let $\phi(\omega)$ be the Fourier transform of $F(t)$:

$$\phi(\omega) = \int_{-\infty}^{+\infty} F(t)e^{-i\omega t}dt \qquad (5)$$

Then:

$$\phi(\omega) = p\varphi(\omega) + \cdots + p(1-p)^k\varphi(\omega)^{k+1} + \cdots \qquad (6)$$

$$|\varphi(\omega)| \leq 1, \ \varphi(0) = 1 \qquad (7)$$

because $\varphi(\omega)$ is a Fourier transform of a probability distribution density. The geometrical progression series in (6) converges to:

$$ɸ(\omega) = \frac{p\varphi(\omega)}{1-(1-p)\varphi(\omega)} \tag{8}$$

Using (8) one gets:

$$F(t) = \int_{-\infty}^{+\infty} \frac{d\omega}{2\pi} e^{i\omega t} \frac{p\varphi(\omega)}{1-(1-p)\varphi(\omega)} \tag{9}$$

From (8) also follows that:

$$\varphi(\omega) = \frac{\frac{1}{p}ɸ(\omega)}{1-(1-\frac{1}{p})ɸ(\omega)} \tag{10}$$

We can now analyze the physical meaning of these mathematical results. Expression (8) gives a conform transformation of the complex plane of $\varphi(\omega)$ onto the complex plane of $ɸ(\omega)$. Due to (7) the physical interest has a transform of a unit cycle which contains all possible Fourier transforms $\varphi(\omega)$ of all possible probability density distributions $f(t)$. This unit cycle is transformed into some part of a unit cycle

$$|ɸ(\omega)| \leq 1 \tag{11}$$

which contains all possible Fourier transforms $ɸ(\omega)$ of all possible probability density distributions $F(t)$ (see fig.1). The unit cycles coincide only in a trivial case $p = 1$. From this follows that for $p < 1$ some probability density distributions $F(t)$ may exist that have no corresponding probability density distributions $f(t)$. We call such probability density distributions "nonclassical" ones. The criterium for $F(t)$ to be nonclassical is to have some $ɸ(\omega)$ outside the "classical" region of the unit cycle shown in fig1. This obviously will be the case for $p < 1$ for Fourier components having amplitude close to 1 and phase close to $\pi$. An example of such $ɸ(\omega)$ can be the Fourier transform of a probability density distribution for the photon antibunching effect, because the deep minimum of $F(t)$ at zero delay demands a high frequency Fourier component $ɸ(\omega)$ with amplitude close to $|ɸ(\omega)| = 1$ and phase close to $\pi$ while in early experiments $p \ll 1$ [2] and in more recent experiments with single atoms in a trap $p < 0.05$ [3]. Due to our consideration, overlooked with the probability $1 - p$ events previous to the photon detection at time $t$ should smear out the deep minimum of $F(t)$ at zero delay.

It is often said that photon antibunching effect demonstrates the quantum nature of light, because it cannot be explained with interference of waves. According to our consideration this effect cannot be explained in a particle picture as well. We think the origin of this paradox lies in the fact that the semi-classical theory of light is a bad approximation in principle. Atom in a trap and light should be described with a common state vector. The detection of a photon causes the jump of the atom from the excited to the ground state directly without any propagation effects through the apparatus. In other words the detection of a photon acts back onto the radiating source.

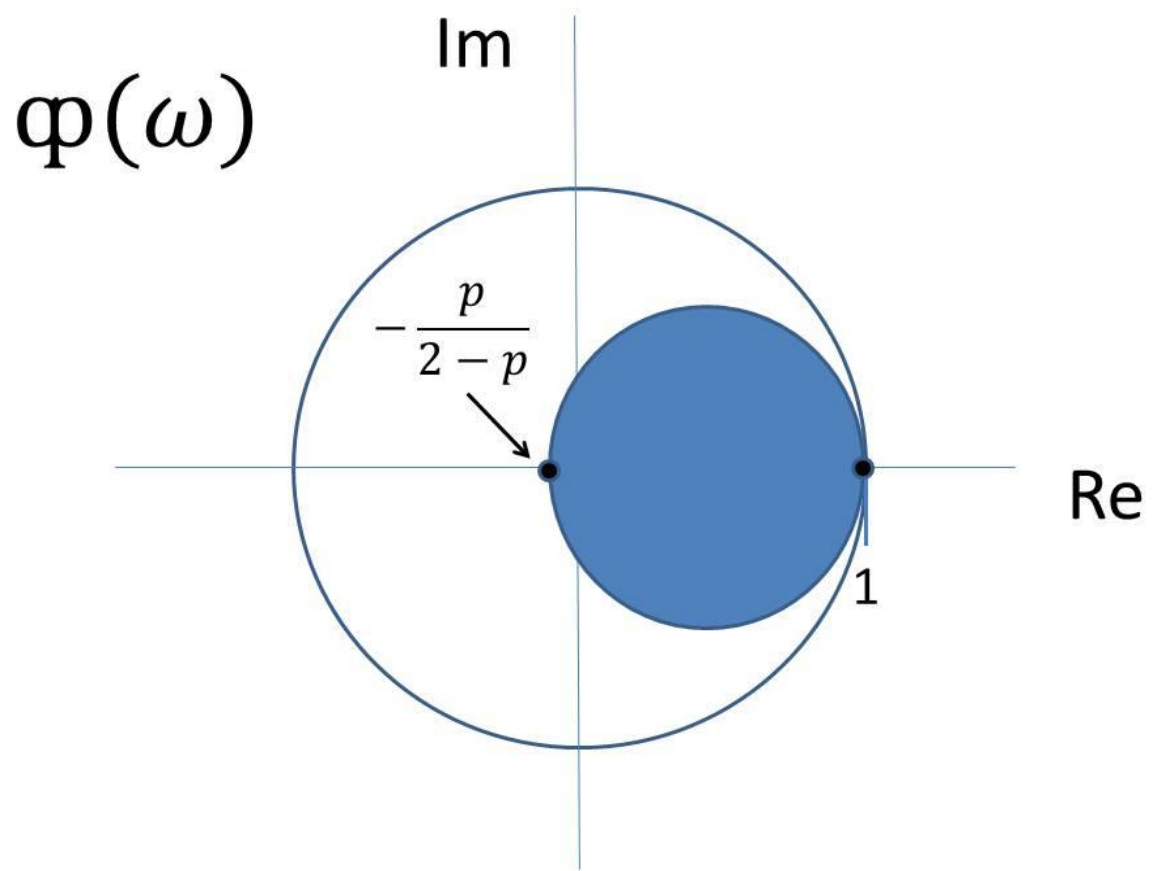

Fig.1. The unit cycle containing all possible Fourier components ϕ($\omega$) of all possible probability density distributions $F(t)$ and the "classical" region (blue on line) of it.